\documentclass[
    ,final            
  ]
  {aipproc}

\layoutstyle{6x9}

\begin{document}

\title{Chandra Localizations of LMXBs: IR Counterparts and their Properties}

\classification{97.80.Jp}
\keywords      {low mass X-ray binaries, infrared counterparts}

\author{Stefanie Wachter}{
  address={Spitzer Science Center, Caltech}
}

\author{Joseph W. Wellhouse}{
  address={Harvey Mudd College}
}

\author{Reba M. Bandyopadhyay}{
  address={Oxford University}
}

\begin{abstract}

We present new Chandra observations of the low mass X-ray binaries (LMXBs) 
X1624$-$490, X1702$-$429, and X1715$-$321 and the search for their 
Infrared (IR) counterparts. We also report on early results from our 
dedicated IR survey of LMXBs.
The goal of this program is to investigate whether IR counterparts 
can be identified through unique IR colors and to trace the origin of 
the IR emission in these systems.
\end{abstract}

\maketitle


\subsection{Infrared Properties of LMXBs}

Traditionally, LMXBs have been studied in the optical and UV part of 
the spectrum. In order to explore the IR properties of LMXBs and to 
investigate the most heavily absorbed sources in the Galactic Bulge, 
we are undertaking a dedicated IR survey of all LMXBs. In addition 
to our own observations, we have also searched the literature for 
published IR magnitudes for these sources. For the
brightest LMXBs in fields with moderate crowding, we extracted $J$, $H$, 
and $K$ magnitudes from the 2MASS database. Selected early results from 
our survey are summarized in Table~1 below. Most of the observations 
were obtained with the 1.5m telescope at CTIO. Photometry was performed 
with DAOPHOT II and standardized
magnitudes were derived through comparison with 2MASS.

Figure 1 shows the position of the individual LMXBs in the IR 
color-color diagram (filled circles). Open circles indicate multiple 
measurements  of the
same sources. Also shown are the main sequence and giant branch tracks. 
The intrinsic variability of the LMXBs limits the predictive power 
of the IR colors (see e.g. Sco X$-$1). A few sources reveal the 
contribution of a giant mass donor.
X1608$-$52 and X1636$-$536 appear to show very unusual colors. 
These are some of the faintest sources we measured and require 
deeper observations to confirm
our photometry.

Figure 2 shows the positions of the individual LMXBs in the IR 
color-magnitude diagram (note that apparent, not absolute, K 
magnitudes are plotted). The symbols are the same as used in Figure 1. 
For comparison, we also include the location
of field stars from a representative Galactic Bulge field (small filled
circles). The 
different branches visible in the color-magnitude diagram 
distinguish different types of stars. 
The first branch, roughly up to $J-K = 1.8$, corresponds to nearby 
main sequence stars, while the clump of stars around $J-K = 2.0-2.5$ 
represents a superposition of giant stars with different values of 
extinction and distance. The LMXBs appear to preferentially cluster 
in an almost vertical strip around $J-K = 0$. GX 13+1 stands out 
as a remarkably red source.


\begin{figure}
  \includegraphics[height=.5\textheight]{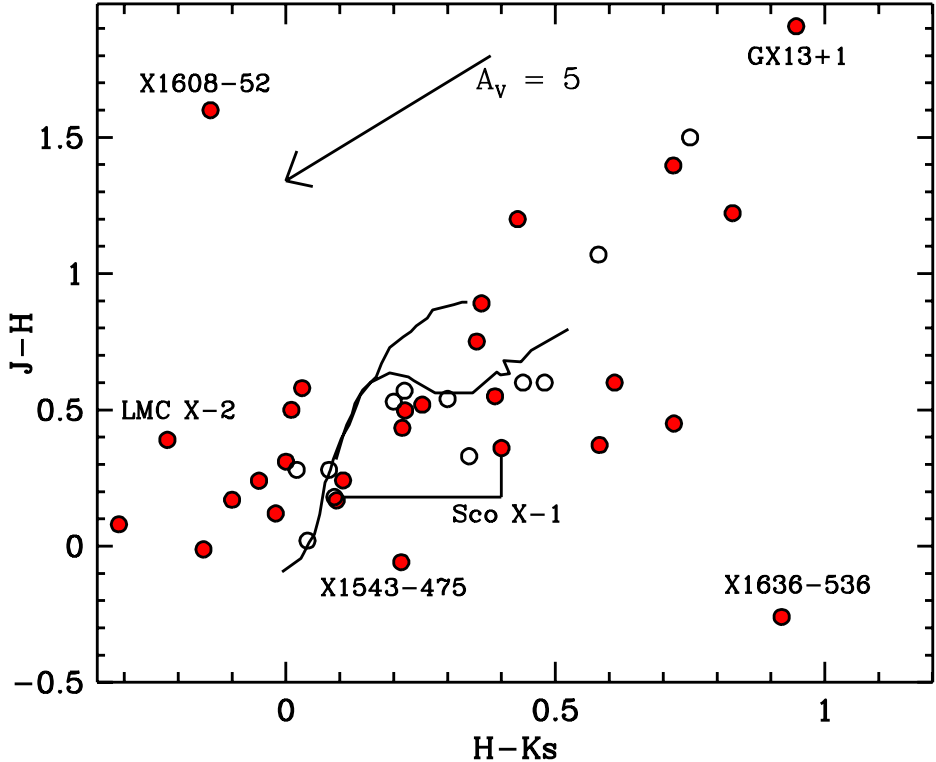}
  \caption{IR color-color diagram for LMXBs. For details, please see text.}
\end{figure}

\begin{figure}
  \includegraphics[height=.5\textheight]{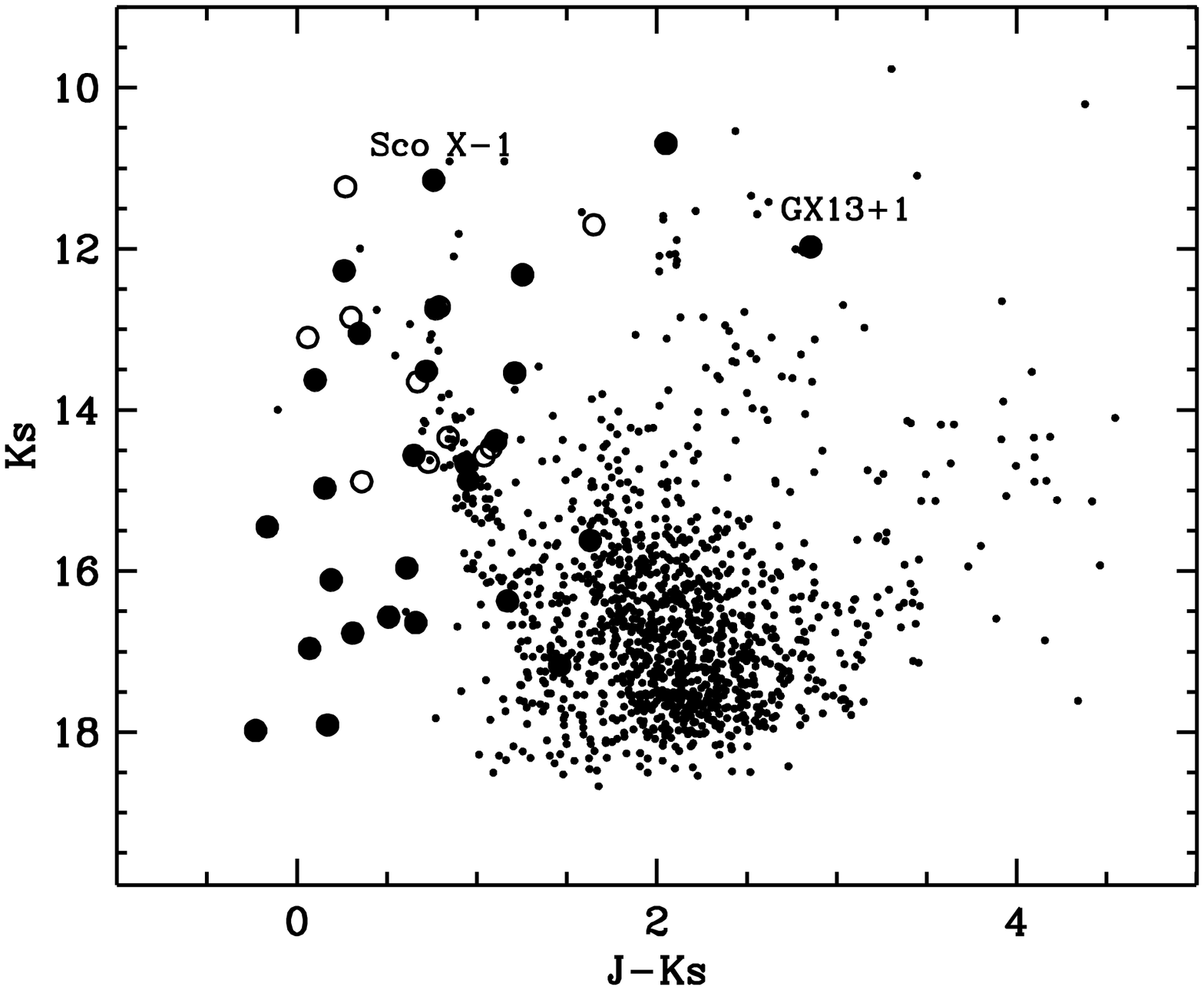}
  \caption{IR color-magnitude diagram for LMXBs. For details, please see text.}
\end{figure}


\begin{table}
\begin{tabular}{crrrrr}
\hline
   \tablehead{1}{c}{b}{Source}
  & \tablehead{1}{r}{b}{$K$}
  & \tablehead{1}{r}{b}{$J-K$}
  & \tablehead{1}{r}{b}{$H-K$}  
  & \tablehead{1}{r}{b}{$J-H$} 
  & \tablehead{1}{r}{b}{Ref.} \\
\hline
 LMC X$-$2   & 17.910 &  0.170 & $-$0.220 &  0.390 & this work \\
 X0614+091   & 16.370 &  1.170 &  0.720 &  0.450 & this work \\
 X0620$-$003 & 14.383 &  1.105 &  0.354 &  0.751 & 2MASS \\
             & 14.470 &  1.080 &  0.480 &  0.600 & this work\\
 X0748$-$676 & 16.960 &  0.070 & $-$0.100 &  0.170 & this work \\
 X0921$-$630 & 13.518 &  0.719 &  0.221 &  0.498 & 2MASS \\
             & 13.650 &  0.670 &  0.340 &  0.330 & Lit. \\
 Cen X$-$4   & 14.663 &  0.938 &  0.388 &  0.550 & 2MASS\\
             & 14.570 &  1.040 &  0.440 &  0.600 & this work\\
 Cir X$-$1   & 10.692 &  2.051 &  0.829 &  1.222 & 2MASS\\
             & 11.700 &  1.650 &  0.580 &  1.070 & Lit. \\
 X1543$-$475 & 14.890 &  0.360 &  0.080 &  0.280 & this work\\
             & 14.970 &  0.155 &  0.214 & $-$0.059 & 2MASS\\
 X1550$-$564 & 15.620 &  1.630 &  0.430 &  1.200  & this work\\
 X1556$-$605 & 17.980 & $-$0.230 & $-$0.310 &  0.080  & this work\\
 X1608$-$522 & 17.170 &  1.460 & $-$0.140 &  1.600  & this work\\
 Sco X$-$1   & 11.147 &  0.760 &  0.400 &  0.360  & 2MASS\\
             & 11.230 &  0.270 &  0.090 &  0.180  & this work\\
 X1636$-$536 & 16.640 &  0.660 &  0.920 & $-$0.260  & this work\\
 X1655$-$40  & 12.744 &  0.772 &  0.253 &  0.519  & 2MASS\\
             & 12.720 &  0.790 &  0.220 &  0.570  & this work\\
 Her X$-$1   & 13.628 &  0.101 & $-$0.019 &  0.120  & 2MASS\\
             & 13.100 &  0.060 &  0.040 &  0.020  & this work\\
 X1658$-$298 & 16.570 &  0.510 &  0.010 &  0.500  & this work\\
 GX 349+2    & 14.563 &  0.650 &  0.216 &  0.434  & 2MASS\\
             & 14.650 &  0.730 &  0.200 &  0.530  & this work\\
             & 14.340 &  0.840 &  0.300 &  0.540  & this work\\
  GX 9+9     & 16.110 &  0.190 & $-$0.050 &  0.240  & this work\\
  GX 1+4     &  7.979 &  2.116 &  0.719 &  1.397  & 2MASS\\
             &  8.060 &  2.250 &  0.750 &  1.500  & Lit. \\
 X1735$-$444 & 16.770 &  0.310 &  0.000 &  0.310  & this work\\
 GX 5$-$1    & 13.540 &  1.210 &  0.610 &  0.600  & Lit. \\
 GX 13+1     & 11.974 &  2.855 &  0.947 &  1.908  & 2MASS\\
 J1819.3-2525 & 12.270 &  0.262 &  0.094 &  0.168  & 2MASS\\
              & 12.850 &  0.300 &  0.020 &  0.280  & this work\\
 X1822$-$371 & 15.450 & $-$0.165 & $-$0.153 & $-$0.012  & 2MASS\\
 Aql X$-$1   & 15.960 &  0.610 &  0.030 &  0.580  & this work\\
 X2023+338   & 12.321 &  1.254 &  0.363 &  0.891  & 2MASS\\
 X2129+470   & 14.873 &  0.953 &  0.582 &  0.371  & 2MASS\\
 Cyg X$-$2   & 13.049 &  0.347 &  0.106 &  0.241  & 2MASS\\
\hline
\end{tabular}
\caption{IR Observations of LMXBs}
\label{tab:a}
\end{table}

\subsection{Chandra Localizations}

\subsubsection{X1624$-$490 and X1702$-$429}


We observed X1624$-$490 on 2002 May 30 and X1702$-$429 on 2003 June 19 with
the Chandra HRC-I for 1 ksec each. 
In the X1624$-$490 data set, a single bright source is detected at the center of the 
30'$\times$30' field. The best position is 
16:28:02.825 $-49$:11:54.61 (J2000) with the
nominal 0.6" positional uncertainty.
In the X1702$-$429 data set, the X-ray binary is the only source detected.
Our best localization 
gives 17:06:15.314 $-43$:02:08.69 (J2000). 

We also obtained deep Ks band observations of each source at the ESO NTT with SOFI and the 
CTIO 4m telescope with ISPI, respectively. A single, faint ($Ks=18.3 \pm 0.1$) source 
is visible inside the Chandra error circle of 
X1624$-$490, and we
propose this source as its IR counterpart. For X1702$-$429, a $Ks= 16.5 \pm 0.07$ source 
is visible at the edge 
of the Chandra error circle. The brightness of both counterpart candidates is comparable to
that of other low mass X-ray binary IR counterparts when corrected for extinction and distance.
For details, please refer to Wachter et al. 2005, ApJ, in press. 

\subsubsection{X1715$-$321}
X1715$-$321 is a poorly studied burster and transient at a distance 
of 5-7 kpc. We obtained a 1 ksec HRC-I observation of the source in 
an effort to detect its quiescent counterpart. No source was 
detected in the observation, placing an
upper limit of $2.8 \times 10^{-14}$ ergs cm$^{-2}$ s$^{-1}$ 
for the quiescent flux from this source.


\begin{theacknowledgments}
The research described in these pproceedings was carried out, in part, 
at the Jet Propulsion Laboratory, California Institute of Technology, 
and was sponsored by the National Aeronautics and Space Administration. 
We made use of data products from
the 2 Micron All Sky Survey, which is a joint project of the 
University of Massachusetts and the Infrared Processing and 
Analysis Center/California Institute of Technology, funded by the 
National Aeronautics and Space Administration and
the National Science Foundation. It also utilized NASA's Astrophysics Data
System Abstract Service and the SIMBAD database operated by CDS, 
Strasbourg, France. SW was supported by Chandra award GO2-3044X. 
SW and JW acknowledge support through Chandra grant GO3-4036X.
\end{theacknowledgments}






\end{document}